\documentclass{nature}
\usepackage{amssymb}
\usepackage{aas_macros}
\usepackage{graphicx}
\usepackage{pdflscape}
\usepackage{hyperref}
\usepackage{txfonts}
\usepackage{textcomp}
\usepackage{threeparttable}
\usepackage{marvosym}

\usepackage{amssymb,amsmath}
\usepackage{xcolor}
\usepackage{ulem}
\usepackage{lineno}
\usepackage{pdfpages}
\usepackage[caption = false]{subfig}
\usepackage{caption}
\makeatletter
\usepackage[symbol]{footmisc}
\newcommand{\EXTTAB}[1] {Supplementary Table}
\usepackage{longtable}

\let\saved@includegraphics\includegraphics
\AtBeginDocument{\let\includegraphics\saved@includegraphics}
\makeatother

\title{The magnetar model's energy crisis for a prolific repeating fast radio burst source}
\begin{document}
\maketitle
\author{
J.~S. Zhang$^{1,2}$\href{https://orcid.org/0009-0005-8586-3001}{\includegraphics[scale=0.08]{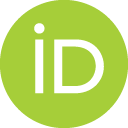}}\footnotemark[1],
T.~C. Wang$^{3,4}$\href{https://orcid.org/0000-0002-9332-5562}{\includegraphics[scale=0.08]{ORCIDiD.png}}\footnotemark[1],
P. Wang$^{1,4,5}$\href{https://orcid.org/0000-0002-3386-7159}{\includegraphics[scale=0.08]{ORCIDiD.png}}\textsuperscript{\Letter},
Q. Wu$^6$\href{https://orcid.org/0000-0001-6021-5933}{\includegraphics[scale=0.08]{ORCIDiD.png}},
D. Li$^{7,1,5}$\href{https://orcid.org/0000-0003-3010-7661}{\includegraphics[scale=0.08]{ORCIDiD.png}}\textsuperscript{\Letter},
W.~W. Zhu$^{1,4,5}$\href{https://orcid.org/0000-0001-5105-4058}{\includegraphics[scale=0.08]{ORCIDiD.png}}\textsuperscript{\Letter},
B. Zhang$^{8,9,10}$\href{https://orcid.org/0000-0002-9725-2524}{\includegraphics[scale=0.08]{ORCIDiD.png}}\textsuperscript{\Letter},
H. Gao$^{3,4}$\href{https://orcid.org/0000-0003-2516-6288}{\includegraphics[scale=0.08]{ORCIDiD.png}}\textsuperscript{\Letter},
K.~J. Lee$^{11,1,12,13}$\href{https://orcid.org/0000-0002-1435-0883}{\includegraphics[scale=0.08]{ORCIDiD.png}},
J.~L. Han$^{1,2,5}$\href{https://orcid.org/0000-0002-9274-3092}{\includegraphics[scale=0.08]{ORCIDiD.png}},
C.~W. Tsai$^{1,4,2,5}$\href{https://orcid.org/0000-0002-9390-9672}{\includegraphics[scale=0.08]{ORCIDiD.png}},
F.~Y. Wang$^{6,14}$\href{https://orcid.org/0000-0003-4157-7714}{\includegraphics[scale=0.08]{ORCIDiD.png}},
Y.~F. Huang$^{6,14}$\href{https://orcid.org/0000-0001-7199-2906}{\includegraphics[scale=0.08]{ORCIDiD.png}},
Y.~C. Zou$^{15}$\href{https://orcid.org/0000-0002-5400-3261}{\includegraphics[scale=0.08]{ORCIDiD.png}},
D.~K. Zhou$^{16}$\href{https://orcid.org/0000-0002-7420-9988}{\includegraphics[scale=0.08]{ORCIDiD.png}},
W.~J. Lu$^{1,2}$\href{https://orcid.org/0000-0001-5653-3787}{\includegraphics[scale=0.08]{ORCIDiD.png}},
J.~T. Xie$^{17}$\href{https://orcid.org/0000-0001-5649-2591}{\includegraphics[scale=0.08]{ORCIDiD.png}},
J.~H. Fang$^{16}$\href{https://orcid.org/0000-0001-9956-6298}{\includegraphics[scale=0.08]{ORCIDiD.png}},
J.~H. Cao$^{1,2}$\href{https://orcid.org/0009-0000-7501-2215}{\includegraphics[scale=0.08]{ORCIDiD.png}},
C.~C. Miao$^{18}$\href{https://orcid.org/0000-0002-9441-2190}{\includegraphics[scale=0.08]{ORCIDiD.png}},
Y.~H. Zhu$^{1,2}$\href{https://orcid.org/0009-0009-8320-1484}{\includegraphics[scale=0.08]{ORCIDiD.png}},
Y.~C. Chen$^{16}$,
X.~F. Cheng$^{16}$,
Y.~N. Ke$^{16}$,
Y.~K. Zhang$^1$\href{https://orcid.org/0000-0002-8744-3546}{\includegraphics[scale=0.08]{ORCIDiD.png}},
L.~X. Zhang$^{15}$\href{https://orcid.org/0009-0002-3020-9123}{\includegraphics[scale=0.08]{ORCIDiD.png}},
S.~Cao$^{12,2}$,
S.~Y. Tian$^{15}$,
Z.~W. Wu$^1$,
C.~F. Zhang$^1$\href{https://orcid.org/0000-0002-4327-711X}{\includegraphics[scale=0.08]{ORCIDiD.png}},
J.~R. Niu$^1$\href{https://orcid.org/0000-0001-8065-4191}{\includegraphics[scale=0.08]{ORCIDiD.png}},
D.~J. Zhou$^1$\href{https://orcid.org/0000-0002-6423-6106}{\includegraphics[scale=0.08]{ORCIDiD.png}},
S.~L. Xu$^{1,2}$,
B.~J. Wang$^1$\href{https://orcid.org/,0000-0002-9434-4773}{\includegraphics[scale=0.08]{ORCIDiD.png}},
H.~X. Chen$^{16}$,
X.~L. Chen$^1$\href{https://orcid.org/0000-0001-5738-9625}{\includegraphics[scale=0.08]{ORCIDiD.png}},
X.~H. Cui$^{1}$\href{https://orcid.org/0000-0002-6165-0977}{\includegraphics[scale=0.08]{ORCIDiD.png}},
Y. Feng$^{16,19}$\href{https://orcid.org/0000-0002-0475-7479}{\includegraphics[scale=0.08]{ORCIDiD.png}},
E. G\"{u}gercino\u{g}lu$^{1,20}$,
Y.~X. Huang$^{12,2}$\href{https://orcid.org/0000-0001-7199-2906}{\includegraphics[scale=0.08]{ORCIDiD.png}},
% J. Li$^{21,22}$\href{https://orcid.org/0000-0003-1720-9727}{\includegraphics[scale=0.08]{ORCIDiD.png}},
D.~M. Li$^3$,
D.~Z. Li$^{21}$,
Y. Li$^{22}$,
L. Lin$^{3,4}$,
X. H. Liu$^{1,2}$\href{https://orcid.org/0000-0002-2552-7277}{\includegraphics[scale=0.08]{ORCIDiD.png}}, 
R. Luo$^{23}$\href{https://orcid.org/0000-0002-4300-121X}{\includegraphics[scale=0.08]{ORCIDiD.png}},
J.~W. Luo$^{24,25}$\href{https://orcid.org/0000-0002-9642-9682}{\includegraphics[scale=0.08]{ORCIDiD.png}},
C.~H. Niu$^{26}$\href{https://orcid.org/0000-0001-6651-7799}{\includegraphics[scale=0.08]{ORCIDiD.png}},
Q.~Y. Qu$^{1,2}$\href{https://orcid.org/0009-0005-5413-7664}{\includegraphics[scale=0.08]{ORCIDiD.png}},
Y.~H. Qu$^{9,10}$\href{https://orcid.org/0000-0003-4721-4869}{\includegraphics[scale=0.08]{ORCIDiD.png}},
H.~M. Tedila$^1$,
C.~J. Wang$^{28}$,
W.~Y. Wang$^2$\href{https://orcid.org/0000-0001-9036-8543}{\includegraphics[scale=0.08]{ORCIDiD.png}},
Y.~B. Wang$^{28}$,
Y.~D. Wang$^{1,2}$\href{https://orcid.org/0000-0002-7372-4160}{\includegraphics[scale=0.08]{ORCIDiD.png}},
S.~M. Weng$^{29, 30}$\href{https://orcid.org/0000-0001-7746-9462}{\includegraphics[scale=0.08]{ORCIDiD.png}},
Y.~S. Wu$^{28}$,
H. Xu$^1$,
A.~Y. Yang$^{1,5}$\href{https://orcid.org/0000-0003-4546-2623}{\includegraphics[scale=0.08]{ORCIDiD.png}},
Y.~P. Yang$^{31}$\href{https://orcid.org/0000-0001-6374-8313}{\includegraphics[scale=0.08]{ORCIDiD.png}},
S.~H. Yew$^{29, 30}$\href{https://orcid.org/0000-0002-5799-9869}{\includegraphics[scale=0.08]{ORCIDiD.png}},
W.~F. Yu$^{32}$,
L. Zhang$^{1,33}$\href{https://orcid.org/0000-0001-8539-4237}{\includegraphics[scale=0.08]{ORCIDiD.png}},
R.~S. Zhao$^{34}$\href{https://orcid.org/0000-0002-1243-0476}{\includegraphics[scale=0.08]{ORCIDiD.png}}

}
\makeatletter
\def\thanks#1{\protected@xdef\@thanks{\@thanks
        \protect\footnotetext{#1}}}
\makeatother

% \linenumbers
\maketitle
\footnotetext[1]{These authors contributed equally to this work.}
\footnotetext{\Letter~ wangpei@nao.cas.cn; dili@tsinghua.edu.cn; zhuww@nao.cas.cn; bzhang1@hku.hk; gaohe@bnu.edu.cn}
\begin{affiliations}
\item National Astronomical Observatories, Chinese Academy of Sciences, Beijing 100101, China
\item University of Chinese Academy of Sciences, Beijing 100049, China
\item School of Physics and Astronomy, Beijing Normal University, Beijing 100875, China
\item Institute for Frontiers in Astronomy and Astrophysics, Beijing Normal University, Beijing 102206, China
\item State Key Laboratory of Radio Astronomy and Technology, Beijing 100101, China 
\item School of Astronomy and Space Science, Nanjing University, Nanjing 210023, China
\item New Cornerstone Science Laboratory, Department of Astronomy, Tsinghua University, Beijing 100084, China
\item Department of Physics, University of Hong Kong, Pokfulam Road, Hong Kong, China
\item Nevada Center for Astrophysics, University of Nevada, Las Vegas, NV 89154, USA
\item Department of Physics and Astronomy, University of Nevada Las Vegas, Las Vegas, NV 89154, USA
\item Department of Astronomy, Peking University, Beijing 100871, China
\item Yunnan Astronomical Observatories, Chinese Academy of Sciences, Kunming 650216, Yunnan, China
\item Beijing Laser Acceleration Innovation Center, Huairou, Beijing 101400, China
\item Key Laboratory of Modern Astronomy and Astrophysics (Nanjing University), Ministry of Education, Nanjing 210023, China
\item School of Physics, Huazhong University of Science and Technology, Wuhan 430074, China
\item Research Center for Astronomical Computing, Zhejiang Laboratory, Hangzhou 311100, China
\item School of Computer Science and Engineering, Sichuan University of Science and Engineering, Yibin 644000, China
\item College of Physics and Electronic Engineering, Qilu Normal University, Jinan 250200, China
\item Institute for Astronomy, School of Physics, Zhejiang University, Hangzhou 310027, China
\item School of Arts and Science, Qingdao Binhai University, Qingdao 266525, China
\item Department of Astronomy, Tsinghua University, Beijing 100084, China
\item Purple Mountain Observatory, Chinese Academy of Sciences, Nanjing 210023, China
\item Department of Astronomy, School of Physics and Materials Science, Guangzhou University, Guangzhou, China
\item College of Physics and Hebei Key Laboratory of Photophysics Research and Application, Hebei Normal University, Shijiazhuang, Hebei 050024, China
\item Shijiazhuang Key Laboratory of Astronomy and Space Science, Hebei Normal University, Shijiazhuang, Hebei 050024, China
\item Institute of Astrophysics, Central China Normal University, Wuhan 430079, China
\item Department of Physics and Astronomy, University of Nevada, Las Vegas 89154, USA
\item Tencent Youtu Lab, Shanghai 200030, China
\item National Key Laboratory of Dark Matter Physics, School of Physics and Astronomy, Shanghai Jiao Tong University, Shanghai 200240, China
\item Laboratory for Laser Plasmas and Collaborative Innovation Centre of IFSA, Shanghai Jiao Tong University, Shanghai 200240, China
\item South-Western Institute for Astronomy Research, Key Laboratory of Survey Science of Yunnan Province, Yunnan University, Kunming, Yunnan 650500, China
\item Shanghai Astronomical Observatory, Chinese Academy of Sciences, Shanghai 200030, China
\item Centre for Astrophysics and Supercomputing, Swinburne University of Technology, Hawthorn 3122, Australia
\item Guizhou Normal University, Guizhou Provincial Key Laboratory of Radio Astronomy and Data Processing, Guiyang 550001, China

\end{affiliations}
\vspace{0.15in}
\begin{abstract}
% 200 words
Fast radio bursts (FRBs) are widely considered to originate from magnetars that power the explosion through releasing magnetic energy. Active repeating FRBs have been seen to produce hundreds of bursts per hour and can stay active for months, thus may provide stringent constraints on the energy budget of FRBs' central engine.
Within a time span of 214 days, we detected 11,553 bursts  from the hyper-active FRB 20240114A that reached a peak burst rate of 729 hr$^{-1}$. This is the largest burst sample from any single FRB source, exceeding the cumulative total of all published bursts from all known FRBs to date. Assuming typical values of radio efficiency and beaming factor, the estimated total isotropic burst energy of this source exceeds 86\% of the dipolar magnetic energy of a typical magnetar. 
The total released energy from this source exceeds that of other known repeaters by about one and a half orders of magnitude, yielding the most stringent lower limit of $\bf{4.7\times10^{32}}$ G~cm$^3$ for the magnetar's magnetic moment. The source remained active at the end of this observation campaign. Our findings thus require either the FRB's central magnetar engine's possessing exceptionally high emission efficiency or a more powerful compact object than a typical magnetar.

\end{abstract}

\linespread{1.15}
Fast radio bursts (FRBs) are bright, millisecond duration radio transients occurring at cosmological distances\cite{lorimer2007,bailes2022,petroff2022}. 
Although their origins and emission mechanisms remain highly uncertain, the similarities between FRB pulses and Galactic magnetar signals\cite{bochenek2020,chime2020} have led to numerous magnetar models being proposed\cite{kumar2017,lu2020, yangzhang2021, metzger2019, margalit2020}.
A small fraction of FRBs have been observed to repeat, including a few producing hundreds of bursts during their peak hour and adding up to thousands of pulses in one episode\cite{li2021, niu2022, xu2022, zhang2022, zhang2023}. 
Such prolific repeaters require an active central engine and a substantial energy reservoir. The most stringent constraints to date comes from  FRB 20220912A and FRB 20201124A, both released $\sim$2\% of all available magnetic energy of a typical magnetar\cite{xu2022, zhang2022, zhang2023}. 

In response to alerts from the Virtual Observatory Event (VOEvent) Service of the Canadian Hydrogen Intensity Mapping Experiment Fast Radio Burst project (CHIME/FRB VOEvent)\cite{abbott2024}, we conducted extensive L-band (1-1.5 GHz) observations of the hyper-active repeating FRB 20240114A using the Five-hundred-meter Aperture Spherical radio Telescope (FAST)\cite{nan2011, li2018}. 

\begin{figure*}
\centering
\includegraphics[width=1\textwidth]{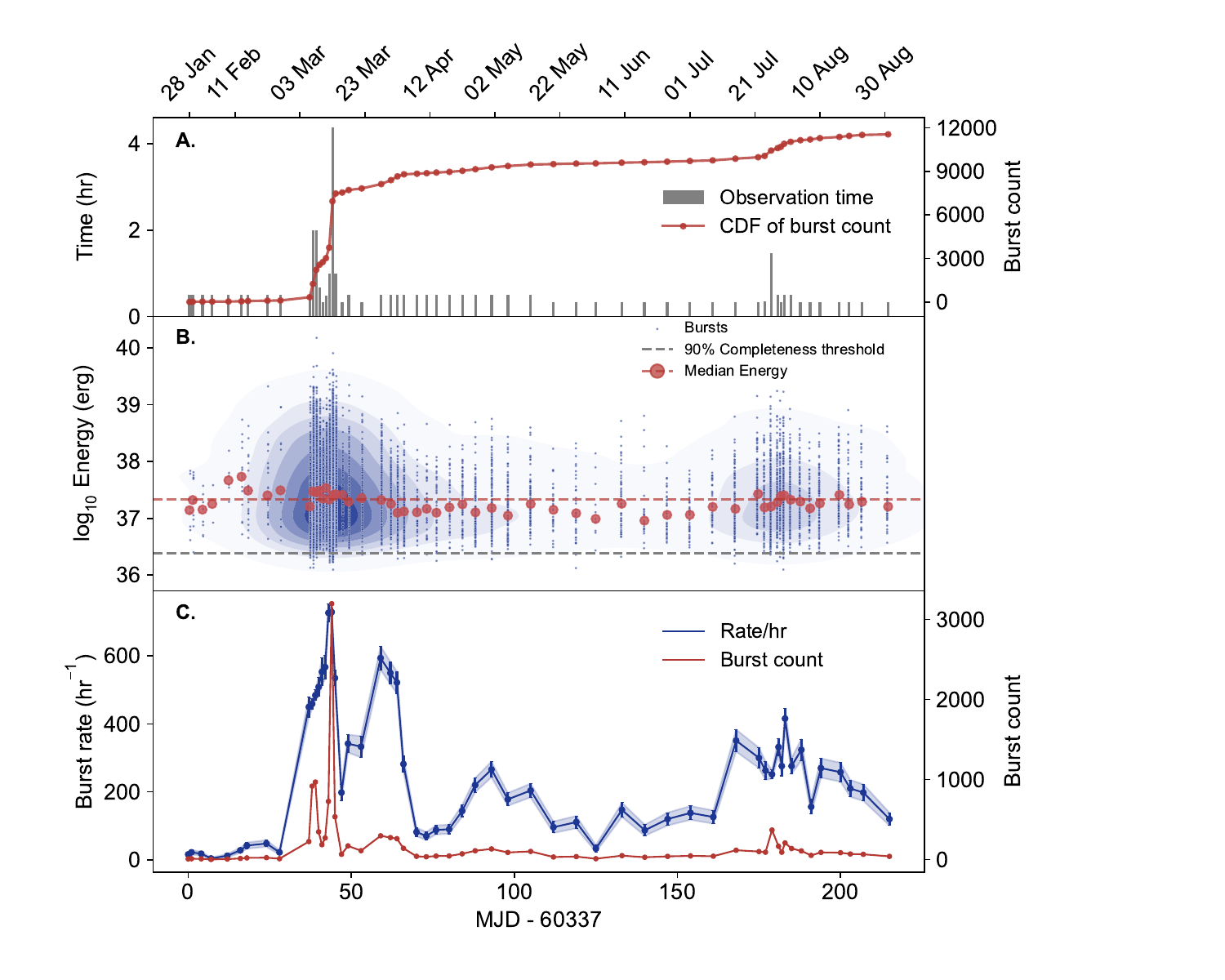}
\caption{{\bf The time-dependent energy distribution and detection of the bursts from FRB 20240114A during the observation campaign.} Panel A: The length of each observation session is represented by the grey bar, while the cumulative number distribution of the bursts is indicated by the red solid line with dots. Panel B: Time-dependent burst energy distribution. The blue dots indicate the bursts, while the red dots represent the daily median energy value. The red dashed line shows the median value of all the bursts. The grey dashed line indicates the 90\% detection completeness threshold of FAST (0.023 Jy ms for an assumed pulse width of 3 ms\cite{li2021}). The blue contour is the 2D kernel density estimation (KDE) of the isotropic burst energies. Panel C: The red colouration denotes the burst counts, and the blue colouration denotes the burst rates with Poisson counting errors, for each observation day.}\label{fig1}
\end{figure*}

A total exposure time of 33.86 hours were spent  between 28 January 2024 and 29 August 2024 (UT), resulting in 11,553 robustly-detected bursts, more than all published bursts from all known FRBs\footnote{see Refs.~\cite{xu2023, fang2025}, \url{https://blinkverse.zero2x.org/} and \url{https://www.wis-tns.org/}}, above the 0.026~Jy~ms ($\sim$12$\sigma$) fluence threshold. 
The observation dates, session length (between 0.3 and 4 hrs), bursts counts and bursts rates   are illustrated in Fig.\ref{fig1}. 
The source displayed persistent activity throughout the  campaign, with a  variable burst rate spanning from 4 hr$^{-1}$ to 729 hr$^{-1}$. 
The burst rate demonstrated two distinct episodes, with a primary peak on 12 March (UT) and a secondary peak on 28 July 2024 (UT).  
The temporal patterns are visualized in the energy-epoch kernel density estimation (KDE) color map (Fig.~\ref{fig1}B). 
During six observation sessions, the burst rate surpassed the previously reported maximum of 542 hr$^{-1}$ from FRB 20201124A 
\cite{zhang2022}. 
These observations confirm FRB 20240114A as an exceptionally active FRB.

We measured the peak flux density, pulse width, bandwidth, central frequency, fluence, and isotropic equivalent energy of each burst. 
The structure-optimized observational dispersion measure (DM) \cite{nimmo2021, lin2023} was also determined, yielding a median value of 529.1~$\mathrm{pc\ cm^{-3}}$ with a standard deviation of 1.3~$\mathrm{pc\ cm^{-3}}$. 
No significant trend was observed in the temporal evolution of the median value of these parameters. 
We carried out a comprehensive correlation analysis among the measured parameter pairs of the bursts. 
The Pearson correlation coefficient (PCC) indicates an absence of significant correlation between the parameter pairs, except for those that are inherently interdependent (see Methods). 
We tested whether high burst rates could cause systematic feedback on the FRB and its propagation properties, as high radiation pressure may affect the surrounding environment\cite{lu2020a}. 
No correlation was found between these observational parameters and the burst rate (see Methods). 
This result suggests that frequent triggering of the bursts has no significant impact on the circumburst environment and the radiation mechanism. 

The waiting time distribution are clearly bimodal, featuring  two log-normal peaking around 7.11 seconds and 34 milliseconds, respectively. 
Compared with other active repeaters' waiting time distributions, such tens of millisecond shorter-time-scale peak has been commonly seen, while the longer-time-scale peak reflect the activity level of the particular source (see Methods). The persistent millisecond scale waiting time peak may reflect the characteristic time scale of the energy release processes.

\begin{figure*}
\centering
\includegraphics[width=0.7\textwidth]{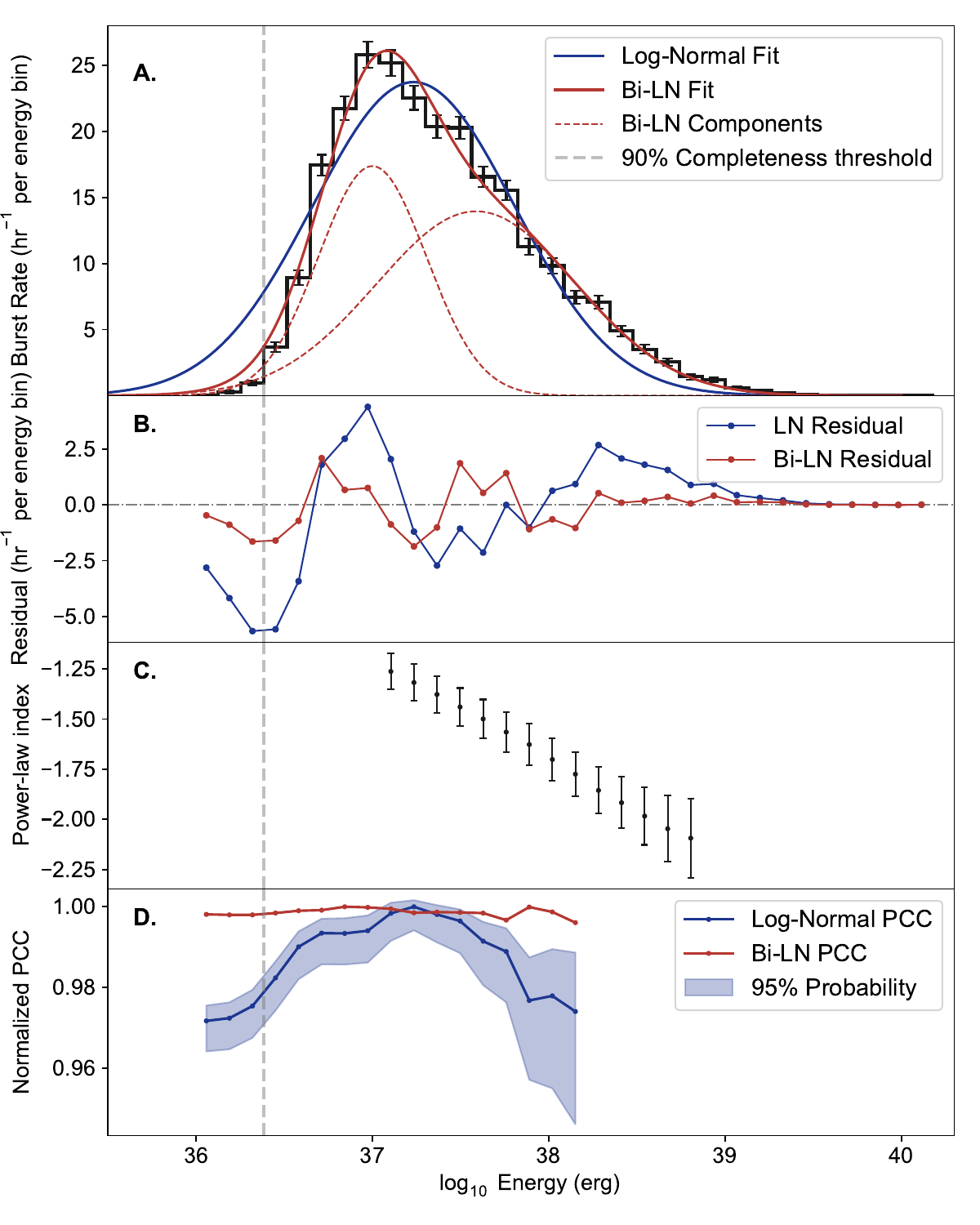}
\caption{{\bf Burst rate distribution of the isotropic equivalent energy for bursts from FRB 20240114A.} The 90\% detection completeness threshold is shown by the vertical gray dashed line. 
Panel A: Burst rate distribution of energy. The blue solid line represents the best-fit single log-normal model of the distribution. The red solid line indicates the optimal bimodal fit, with its individual components delineated as red dashed lines.
Panel B: The residuals of the single log-normal function and bimodal fits to the burst rate distribution. The blue line corresponds to the residuals from the single log-normal function fit, while the red line shows the residuals from the bimodal fit.
Panel C: The spectral index of single power-law fitting as a function of the energy threshold.
Panel D: The normalized Pearson Correlation Coefficient (PCC) as a function of the energy threshold. The blue solid line denotes the results of the log-normal (single log-normal function) fit, and the blue shaded region represents the 95\% probability interval of the normalized PCC. The red solid line indicates the normalized PCC obtained from a bimodal fit using two log-normal components.
}\label{fig2}
\end{figure*}

The burst energy distribution of FRB20240114A (Fig. \ref{fig2}A) exhibits a characteristic energy $E_c \sim 10^{37}$ erg, below which the burst rate drops robustly. Similar characteristic energy were also reported in other active repeaters\cite{li2021, niu2022, zhang2022, zhang2023}. Power-law fits applied separately above different energy thresholds yield non-converging indices (Fig.~\ref{fig2}C). This indicates that a power-law function does not adequately describe the distribution, even above the characteristic energy. We then modeled the distribution using single and bimodal log-normal functions. The single log-normal model shows significant systematic deviations in both the low- and high-energy regions (Fig.~\ref{fig2}B, D). Statistical metrics ($\bar{R}^2$, reduced $\chi^2$, AIC) confirm that the bimodal function fits better represent the data (see Methods).
This complex bimodal distribution suggests the presence of diverse burst types, indicating multiple burst populations or distinct emission mechanisms in repeating FRBs\cite{petroff2022}.

\begin{figure*}
\centering
\includegraphics[width=1\textwidth]{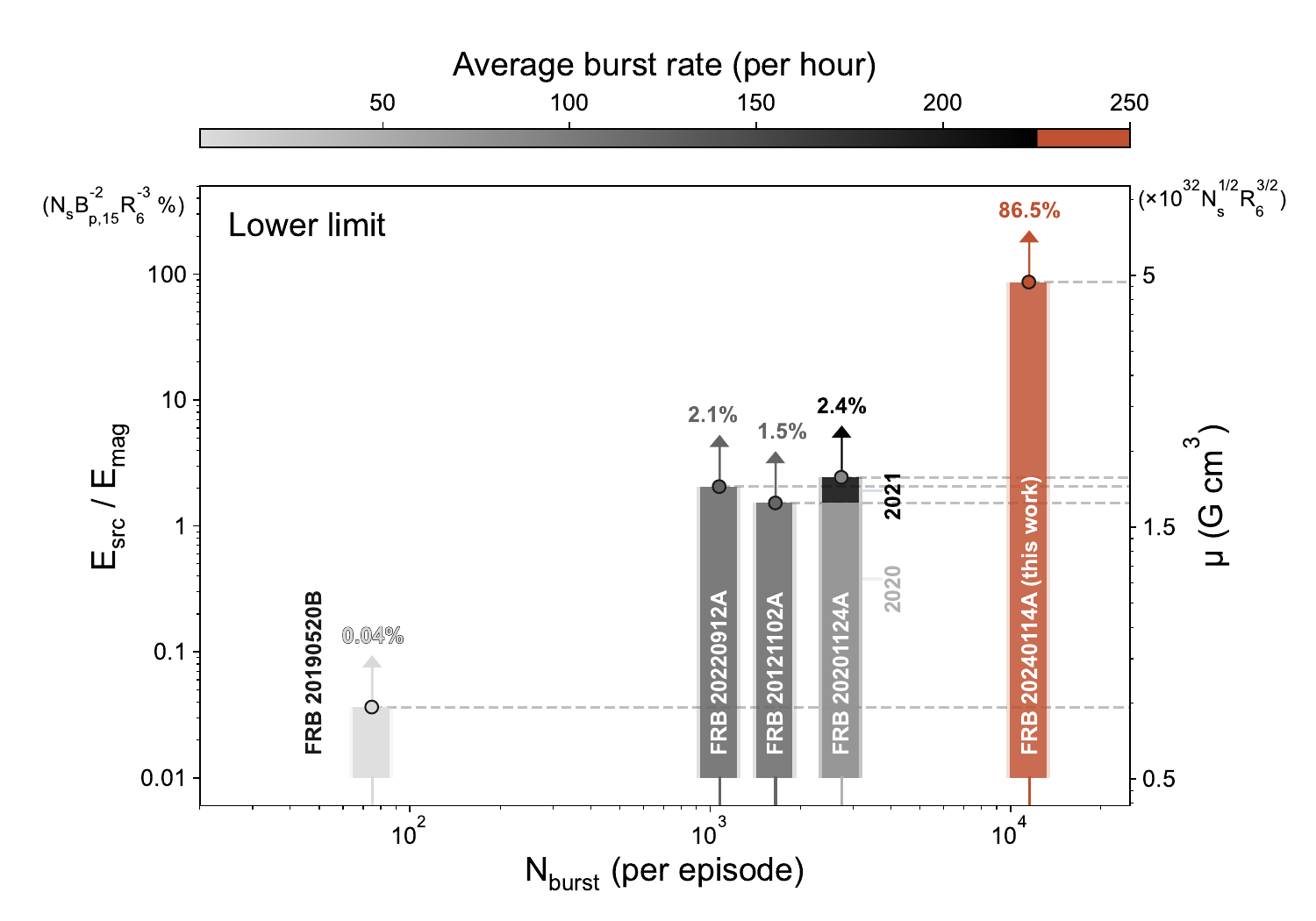}
\caption{{\bf The isotropic energy fraction and magnetic moment of FRBs.} The plot presents the ratio between isotropic equivalent energy of FRBs and the dipole magnetic energy of a typical magnetar, given as $E_{\text{src}}/E_{\text{mag}}$, alongside the burst number per episode, $N_{\text{burst}}$, for active repeating FRBs. The percentages indicated on the bars represent the corresponding ratio for each FRB source. The lower limit on the magnetic moment of the central magnetar is shown on the right axis. Both the ratio and magnetic moment are normalized by $N_s$, reflecting different parameter assumptions. The colorbar shows the average burst rate during each episode.}\label{fig3}
\end{figure*}

The substantial energy emitted by FRB~20240114A imposes severe constraints on the energy budget of magnetar models.
We infer from observation the ratio between the total isotropic equivalent energy released ($E_{\rm src}$) and the magnetar's entire dipole magnetic energy reservoir ($E_{\rm mag}$) as
\begin{equation}
    \frac{E_{\rm src}}{E_{\rm mag}} 
    = \frac{E_{\rm tot}N_s}{2\mu^2/(3R^3_6)}
    = \frac{E_{\rm tot}~ F_{\rm b,0.1}~ \eta_{\rm r,0.0001}^{-1}~ \zeta^{-1}}{1/6B^2_{p,15}R_6^3}\ \approx 86.5\%N_s(\frac{B_p}{10^{15}~\mathrm{G}})^{-2}(\frac{R}{10^{6}~\mathrm{cm}})^{-3},
\end{equation} 
where $E_{\rm tot}$ is the total observed radio energy of the bursts, $N_s$ represents emission properties normalized to typical values, $\mu$ is the magnetic moment of the central magnetar,
$\zeta = 0.0066$ is the observational duty cycle used to scale emission from unobserved periods,
$\eta_{\rm 0.0001}$ is the radio radiation efficiency in units of $10^{-4}$,
$F_{\rm b,0.1}$ is the global beaming factor in units of $0.1$, while
$B_{p}$ and $R$ are the polar surface magnetic field strength and radius normalized with a typical central magnetar, respectively (see Methods).
To satisfy the energy budget, the magnetar must have a polar magnetic field strength of $B_p > 9.4 \times 10^{14} N_s^{1/2} R_6^{-3/2}~\mathrm{G}$, corresponding to a magnetic moment of $\mu > 4.7\times10^{32} N_s^{1/2}R^{3/2}_6~\mathrm{G~cm^3}$, where $R_6$ is the magnetar radius in units of $10^6$ cm.
Fig.\ref{fig3} illustrates the ratio between $E_{\rm src}$ and $E_{\rm mag}$, as well as the lower limit on the central magnetar’s magnetic moment for this source and other active repeaters\cite{li2021, niu2022, xu2022, zhang2022, zhang2023}. The total energy of this source exceeds that of other active repeaters by more than a factor of 35, approaching the total available magnetic energy of a  magnetar. This result imposes the most stringent constraint to date on the FRB central engine model from an energy budget perspective. 

There are two main categories of magnetar models: one involving emission from within the magnetosphere\cite{kumar2017,lu2020, yangzhang2021}, and another from relativistic shocks outside the light cylinder\cite{metzger2019, margalit2020}. 
Recent observations of polarization and scintillation properties favor the magnetospheric models\cite{luo2020, niu2024, mckinven2025, nimmo2025, jiang2025}.
The extreme and continuous activity of this prolific source indicates a substantial energy emission and frequent triggering, leading to a crisis of those magnetar models that invoke a large global beaming factor and low radiation efficiency. 
The synchrotron maser shock model\cite{metzger2019, margalit2020}, in particular, typically involving a low radio efficiency and a wide beaming angle, is thus strongly disfavored. 
The magnetospheric models\cite{kumar2017,lu2020, yangzhang2021}, which invoke narrow emission beams and flexible radio efficiencies\cite{szary2014,cook2024,quzhang2024}, may still be possible, even though contrived conditions are required to satisfy observational constraints.
The extremely high event rate also poses a challenge to the FRB triggering mechanism. Starquakes, which are considered the leading trigger scenario for FRBs, may still be able to account for the observed average burst rate (249 hr$^{-1}$) during our 214-day observation campaign, according to the estimation in Ref.\cite{wangwy2024}. 
The high event rates may also cause distortions in the magnetic field lines, thus challenging the possibility of generating coherent emission at millisecond-level waiting timescales based on our observations.

FRB~20240114A remained active toward the end of our observation campaign reported in this work. Continued monitoring will  place even more stringent constraints on the nature of its central engine, potentially further focusing the magnetar models discussed above. Given the persistent and highly active behavior exhibited by FRB~20240114A over a time span of more than 8 months, the central engine of this source must be characterized by either an exceptionally high radio emission efficiency, or a much more energetic compact object than is conventionally anticipated in prevailing isolated magnetar models.

\bibliographystyle{naturemag}

% \bibliography{method}

\begin{addendum}
 \item This work is supported by National Natural Science Foundation of China (NSFC) Programs Nos. 12588202, 11690024, 11725313, 11988101, 12041303, 12041306, 12203045, 12233002, 12303042, 12403100, 12421003, 12447115, U1731238, U2031117, W2442001; the CAS International Partnership Program No. 114-A11KYSB20160008; the CAS Strategic Priority Research Program No. XDB23000000; the National Key R\&D Program of China (Nos. 2017YFA0402600, 021YFA0718500), the National SKA Program of China (Nos. 2020SKA0120200, 2020SKA0120300, 2022SKA0130104) and the China Postdoctoral Science Foundation (CPSF) under Grant Nos. GZB20240308, GZB20250737, 2025T180875, 2025T180875. 
P.W. acknowledges support from the CAS Youth Interdisciplinary Team, the Youth Innovation Promotion Association CAS (id. 2021055), and the Cultivation Project for FAST Scientific Payoff and Research Achievement of CAMS-CAS.
D.L. is a New Cornerstone investigator. 
W.W.Z. is supported by the CAS Project for Young Scientists in Basic Research, YSBR-063. 
Y.F.H. acknowledges the support from the Xinjiang Tianchi Program. 
Y.F. is supported by the Leading Innovation and Entrepreneurship Team of Zhejiang Province of China grant No. 2023R01008, and by Key R\&D Program of Zhejiang grant No. 2024SSYS0012. 
 This work made use of data from FAST, a Chinese national mega-science facility built and operated by the National Astronomical Observatories, Chinese Academy of Sciences. 
 This work made use of the data from FAST FRB Key Science Project. 
 This research made use of the CHIME/FRB VOEvent Service, BlinkVerse and TransientVerse.
\item[Author Contributions] 
J.S.Z. and T.C.W. led the FAST data analysis.
P.W., D.L., H.G., B.Z. and W.W.Z. are the conveners of the project, coordinated the science team, and launched the FAST observational campaign on FRB~20240114A. 
Q.W., D.K.Z., W.J.L., J.T.X., J.H.F., J.H.C., C.C.M., Y.H.Z., Y.C.C., X.F.C., Y.N.K., Y.K.Z., Y.F., S.C., S.Y.T., L.X.Z., Z.W.W., C.F.Z., J.R.N., D.J.Z., S.L.X., B.J.W. participated in the FAST data analysis. 
B.Z. and Y.P.Y. led the theoretical interpretation.
J.S.Z., P.W. and D.L. contributed to the writing of the manuscript.
All authors discussed the contents and form the final version of the paper.
\item[Competing Interests] The authors declare that they have no competing financial interests.
\end{addendum}


\clearpage
\newpage

\begin{methods}

\setcounter{figure}{0}
%\listoffigures
\captionsetup[figure]{labelfont={bf},labelformat={default},labelsep=period,name={Extended Data Fig.}}

\section{Observation Campaign and Data Reduction}
FAST monitored the repeating FRB 20240114A continuously from 2024 January 28 to 2024 August 29 (UT), using the central beam of 19-beam L-band receiver, initially pointing to the coordinate of RA$=21^h27^m39^s.888$, Dec$=+04^\circ21'0''.36$ as reported by CHIME\cite{shin2025}. We then calibrated the pointing to the position of  RA$=21^h27^m39^s.84$, Dec$=+04^\circ19'46''.3$ localized by MeerKAT\cite{tian2024}  starting from 2024 February 15 (UT). During the observation campaign, 57 observation sessions with a total of 33.86 hours exposure time were conducted. The FAST 19-beam L-band receiver covers a frequency bandwidth of 1.0-1.5 GHz \cite{nan2011, li2018}. Most of the data were recorded in 8-bit PSRFIT format over 4096 frequency channels with a time resolution of 49.152 $\mu$s, while some were recorded with different parameters (see Supplementary Table~\ref{tab:obslog}).

We performed the same dedicated single-pulse search described in Ref.\cite{wang2025}, using the PRESTO software\cite{ransom2001}.  We created the de-dispersed time series for each pseudo-pointing over a range of DMs of 515-540 pc cm$^{-3}$, and also applied a zero-DM matched filter to mitigate radio frequency interference (RFI) in blind-search. All the potential candidate plots were subsequently inspected visually.

We also employed the improved deep learning-based search tool DRAFTS\cite{zhang2025} and the AI-aided data filtering pipeline described in Ref.\cite{wang2024} to cross-verify the search results.
Then, we used the equivalent burst duration $W_{\rm eq}$ of each burst, which was computed by dividing the fluence by the peak flux, to refine the detected fluence signal-to-noise ratio (SNR$_{\rm f}$, corresponding to integrated flux and the burst energy). 

A total of 11,553 bursts were detected during the observation campaign above our detection threshold SNR$_{\rm f} >12$ (see Supplementary Table~\ref{tab:bursttab} for the full catalog).

The FAST receiver experiences saturation when the radio flux density reaches levels ranging from hundreds of Jansky to mega-Jansky. To identify potential saturation events in the recorded data streams, we examined all Stokes components in epochs where 50\% of the frequency channels met one of the following conditions: a) the channel is fully saturated (255 value in 8-bit channels), b) the channel is zero-valued, c) the Root Mean Square (RMS) of the bandpass is less than 2. We identified saturation events in our observations conducted from March 8 to 10. Data affected by saturation during these periods were excluded from further analysis.

Since FAST records data in 8-bit format, saturation also occurs when pulse brightness exceeds the dynamic range limit of 255, leading to an underestimation of flux. Pulses affected by saturation and potential flux underestimation were marked in the Supplementary Table~\ref{tab:bursttab}.

\section{Burst Characterization and Calibration}

We refined the measurement of DM value for each burst using the `DM-power' package, which employs an algorithm that optimally weights the pulse structure at each Fourier frequency\cite{lin2023}. 
For bursts with obscured morphological structures in the time-frequency dynamic spectrum or those with low SNR, the DM-power package may introduce overfitting artifacts. 
In such cases, we utilized the daily average DM values instead. 
The median value of the DM is 529.1 $\mathrm{pc~cm^{-3}}$ with a standard deviation of 1.3 $\mathrm{pc~cm^{-3}}$, which is consistent with the CHIME’s measurement\cite{shin2025}. 

The frequency boundaries of the bursts were calculated using a systematically adjusted method based on the behavior of the first derivative of the cumulative distribution function (CDF) of the spectrum\cite{xie2025}. 
For bursts truncated by the upper or lower limits of the observed frequency range, a frequency-extended Gaussian fit was applied to correct the bandwidth\cite{hu2025}. 
In cases where the bursts showed complex morphological structures or were too faint to yield a reliable result, a boxcar bandwidth was adopted instead. 

We injected a 10 K equivalent noise calibration before each session, which was used to scale the data to $T_{\rm sys}$ units. 
The frequency-averaged specific peak flux density $S_\nu$ within the frequency boundaries for each burst was calibrated against the baseline noise level, and then measured for the amount of pulsed flux density above the baseline.  
The temporal variation of the background was reconstructed by the telescopic gain's dependence curve on the zenith angle and observation frequency\cite{jiang2020}. 
This allowed us to convert $S_\nu$ from Kelvin units to Jansky. 
We note that the position of this source was updated on February 15, 2024 (UT); prior to that, telescope pointing had an error of approximately 1 arcmin from the accurate location. 
Consequently, we corrected the flux density of these bursts using a factor derived from the beam shape of FAST's L-band receiver\cite{jiang2020}.

The frequency-averaged specific fluence of each burst was derived by $F_{\nu}$ = $S_{\nu}\times W_{\rm eq}$ in units of ${\rm erg \ cm^{-2} Hz^{-1}}$ or Jy~ms.  It is noted that $F_{\nu}$ was calculated by averaging the pulse signal over the determined frequency range of each burst, rather than the full observing bandwidth. 
This approach ensures that the fluence was derived solely from the emission observed within the identified frequency band. A more detailed analysis of morphological classification and drift rate measurements is presented in a separate paper \cite{zhang2025b}. For further analysis on the polarization properties of this source, including the rotation measure (RM) and Faraday conversion, see the forthcoming paper (T.-C. Wang et al., in preparation). Multi-band radio observations link FRB 20240114A to a variable radio continuum source, classified as a flare radio source (FRS) rather than a persistent radio source (PRS) due to significant flux density variations\cite{zhangx2025}.

We analyzed the temporal evolution of the DM, equivalent burst duration, flux density, bandwidth, and central frequency. Taking into account the uncertainties estimated using the bootstrap method, we found no significant temporal trend in any of these parameters, as shown in Extended Data Fig.~\ref{exfig1}.
We also carried out PCC analysis to investigate the relationship between the burst rate trend and other parameters. None of the parameter pairs displays a significant positive or negative correlation within the 95\% confidence interval. These results suggest that the frequent triggering of this source has little impact on the circumburst environment or the radiation mechanism, as evidenced by the absence of significant variations in the observational parameters.
\begin{figure*}
\centering
\includegraphics[width=0.9\textwidth]{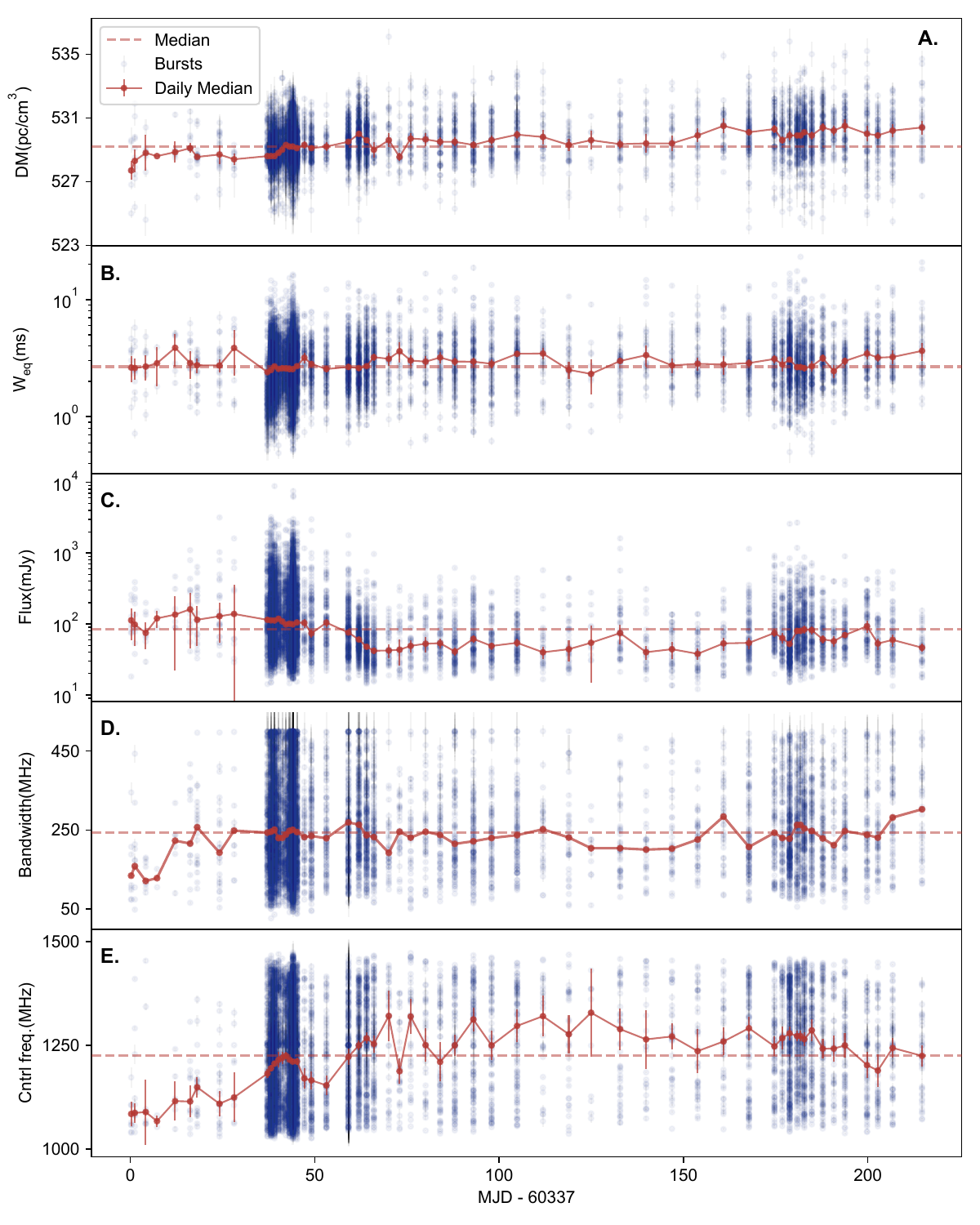}
\caption{{\bf The temporal evolution of FRB 20240114A.} 
Panel A-E shows the DM, $W_{\rm eq}$, flux density, bandwidth and central frequency temporal evolution of FRB 20240114A, respectively. The blue dots denote the corresponding parameter of all the bursts, the red represent the median value of the corresponding parameter in each observation session. The error bars represent the uncertainty in the median value of each parameter. The red dashed lines shows the median value of all bursts for each parameter.}\label{exfig1}
\end{figure*}

\section{Energy Distribution}
Without knowledge of the geometric beaming information of FRBs, we calculate the isotropic equivalent burst energy using the following equation similar to Refs. \cite{zhang2018, gourdji2019}:
\begin{equation}
E = (10^{39} {\rm erg})\frac{4\pi}{1+z} \left(\frac{D_{\rm L}}{10^{28}{\rm cm}}\right)^{2}
\left(\frac{F_{\nu}}{\rm Jy\cdot ms}\right)
\left(\frac{\Delta\nu}{\rm GHz}\right),
\end{equation}
where $D_{\rm L}$ is the luminosity distance, and $\Delta\nu$ is the bandwidth of each burst. 
The optimized estimates depend on the spectral shape of the burst. 
Since the repeating bursts typically have narrow-band emission within the
observing band\cite{zhang2023}, we define $F_{\nu}$ as the average fluence within the frequency range specific to each burst.
Accordingly, the burst energy is obtained by integrating $F_{\nu}$ over $\Delta\nu$, the determined bandwidth of the burst.
This method allows for energy estimation without relying on any assumptions about emission outside the observing band, and thus leads to a more conservative estimate.
The $D_{\rm L}$ of FRB 20240114A is estimated to be 633.87 Mpc based on the standard Planck cosmological model\cite{Planck2016}, corresponding to a spectroscopic redshift of $z = 0.1306$\cite{chen2025}.

The burst rate distribution as a function of energy is weighted by observation time, as the duration of individual observation sessions varies.
We fit the distribution above varying en ergy thresholds, starting from the peak energy, using a single power-law function, as shown in Fig.~\ref{fig2}C. The resulting power-law index displays a decreasing trend with increasing energy threshold, indicating that the PL model does not provide a good fit to the data—even above the characteristic energy.

We then fit the distribution using both LN and Bi-LN model. The residuals of the LN model, shown in Fig.~\ref{fig2}D, reveal systematic structures in both the low- and high-energy regions, indicating that the LN model does not adequately describe the data. To quantify the model performance, we calculate the PCC between the observed and fitted values above varying energy thresholds and normalize it by the maximum PCC of each model. This approach allows us to assess the degree to which each model deviates from the observational data. We estimate the 95\% probability interval of the normalized PCC for the LN model using the Monte Carlo method with 10,000 trials. The normalized PCC for the LN model shows a significant decline on both sides of the characteristic energy, exceeding its 95\% probability interval.

To further evaluate the goodness of each fit, we calculate the adjusted coefficient of determination $\bar{R}^2$ and the reduced chi-squared statistic $\chi^2$ and perform Akaike information criterion (AIC)\cite{burnham2002}
using the following expressions:
\begin{align}
\bar{R}^2 &= 1-\frac{\sum^n_i(N_{\rm obs,i}-N_{\rm mod,i})^2}{\sum^n_i(N_{\rm obs,i} -\bar{N})^2} (\frac{n-1}{n-p-1}), \\
\chi^2 &= \frac{1}{n-p}\sum^{n}_{i=1} \bigg[\frac{(N_{\rm obs,i}-N_{\rm mod,i})^2}{\sigma^2_{\rm i}}\bigg],\\
\mathrm{AIC} &= -2\log \mathcal{L} + 2p,
\end{align}
where $n$ denotes the number of bins, $p$ is the number of the free parameters of each model, $N_{\rm obs,i}$ is the observed value for the $i$th bin, $N_{\rm mod,i}$ is the fitted value for the $i$th bin, $\sigma^2_{\rm i}$ is the error for the $i$th bin, estimated and transferred from the Poisson counting error of the statistical counts of each day, and $\log \mathcal{L}$ is the maximized value of the likelihood function for the model. 
The best-fitting parameters and statistical test results for the two models are listed in Supplementary Table~\ref{tab:fitting}. The adjusted $R^2$, reduced $\chi^2$, and AIC values indicate that the LN model does not adequately describe the data, while the Bi-LN model provides a more plausible representation of the energy distribution for this source. 
The flat tail of the cumulative energy function, analyzed together with data from the Kunming 40-Meter Radio Telescope (KM40M) in the S-band, is discussed in a separate paper \cite{huang2025}.

\section{Energy Budget Constraint}
The isotropic equivalent energy of individual bursts we derived is based on isotropic emission assumption. Coherent radiation from the individual FRB burst generally has a small solid angle $\delta \Omega$. Considering that some bursts may occur in directions that are not observable, we introduce the emission beam as $\Delta \Omega$, and define the global beaming factor as
$F_{\rm b} = \frac{\Delta\Omega}{4\pi}$. This accounts for the possibility that the bursts are confined within a global fan beam, resulting in a net energy that is lower than the isotropic equivalent estimate\cite{zhangb2023}.
Assuming the burst remains active at the average burst rate during the spanning time of our observation campaign, the total source energy should be written as
\begin{equation}
    E_{\rm src} = E_{\rm tot} \times F_{\rm b} \times \eta_{\rm r}^{-1} \times \zeta^{-1}, \label{eq:cal1}
\end{equation}
where $E_{\rm tot}$ denotes the total radio energy of the bursts we observed, $\eta_{\rm r}$ is the radio radiation efficiency, and $\zeta$ is the duty cycle of observations during the observation campaign. 
We adopted a typical value of $\eta_{\rm r} = 10^{-4}$ based on the observational upper limit of $\eta_{\rm r} = 10^{-4} -10^{-5}$, inferred from the peculiar event FRB 20200428D, which was associated with X-ray bursts\cite{mereghetti2020, lic2021}. Furthermore, the global beaming factor, $F_{\rm b,0.1}$, is assigned a typical value of 0.1. 
The duty cycle scaling factor $\zeta$ represents the ratio of the total observation time to the full time span of the campaign, which is used to account for the emission energy of FRB~20240114A during unobserved periods. For the observation campaign described in this work, $\zeta$ is calculated to be $0.0066$. 
We note that FRB 20240114A remained active at the conclusion of the observational campaign. Consequently, the value of $\zeta$ may be subject to change as ongoing observations provide further insights into the source's activity.
Given that $\eta_{\rm r}$ and $F_{\rm b}$ may vary depending on the adopted assumptions and models, we introduce a normalization factor, $N_s = F_{\rm b,0.1} \eta_{\rm r,0.0001}^{-1} (\zeta / 0.0066)^{-1}$, to account for these variations. Here, $\eta_{\rm r,0.0001}$ denotes $\eta_{\rm r}$ expressed in units of $10^{-4}$, and $F_{\rm b,0.1}$ represents $F_{\rm b}$ expressed in units of $0.1$. Assuming $\eta_{\rm r} = 10^{-4}$ and $F_{\rm b} = 0.1$, the total source energy of FRB 20240114A during the 214-day observation period is calculated as:
\begin{equation}
E_{\rm src} = (1.47 \times 10^{47}~\mathrm{erg})N_s.
\end{equation}
Supplementary Table~\ref{tab:energycal} shows the total source energy estimations of this source and other four repeaters using FAST observations reported in Ref.\cite{li2021, niu2022, xu2022, zhang2022, zhang2023}.
It should be noted that the burst energies reported in the literature for FRB 20121102A and FRB 20190520B were calculated using a center frequency of 1.25 GHz, rather than the observing bandwidth or the bandwidth of each burst. Therefore, we have divided the total energies of these two FRBs by 2.5 to allow for comparison with the other FRBs.
The $E_{\rm src}$ of this source corresponds to $\sim 86.5\%N_sB^{-2}_{p,15}R^{-3}_6$ of the total dipolar magnetic energy, given by $E_{\rm mag}=(1/6)B_p^2R^3\simeq (1.7 \times 10^{47} \mathrm{erg})B^2_{p,15}R^3_6$, where the polar surface magnetic field and radius have been normalized using typical magnetar values of $B_p=10^{15}$ G and $R=10^6$ cm, see Ref.\cite{kaspi2017}. 
The magnetic moment of a magnetar is related to its polar surface magnetic field and radius by the relation $\mu=\frac{1}{2}B_p R^3$. 
Here we derive the lower limit of $\mu$ as $ 4.7\times10^{32}N_s^{1/2}R^{3/2}_6~\mathrm{G~cm^3}$, which corresponds to a polar surface magnetic field of $B_p > 9.4 \times 10^{14} N_s^{1/2} R_6^{-3/2}~\mathrm{G}$.
A central magnetar with a magnetic moment below this threshold would be unable to supply sufficient energy to power the source with magnetic energy alone. 

It is the most stringent constraint on the available magnetar energy and magnetic moment, suggesting that the dipolar magnetic energy of the magnetar would be completely depleted in only $\sim$250 days in total if the radio efficiency is indeed as low as $10^{-4}$. Such a stringent energy budget constraint poses significant challenges to certain magnetar models, particularly low-efficiency models involving relativistic shocks, as well as models with large beaming factors that rely on mechanisms triggered in various regions of the magnetar.

\section{Waiting Time}
The burst arrival times were measured at the half-fluence position of each burst. We then converted the site arrival times to the barycentric coordinate time (TCB) at the infinite frequency using the software package TEMPO2\cite{hobbs2006}. 
We calculated the waiting time between two adjacent bursts as $\delta$t = $t_{\rm i+1}-t_{\rm i}$, where $t_{\rm i+1}$ and $t_{\rm i}$ are the barycentered arrival times for the ($i+1$)th and ($i$)th bursts, respectively. All waiting times were calculated for bursts within the same observing session to avoid long gaps of days.
In Extended Data Fig.~\ref{exfig2}a, the waiting time distribution are clearly bimodal, comprising of two log-normal peaks around 7.11 seconds and 34 milliseconds. The bimodal log-normal function provides a well overall fit, with $\bar{R}^2 = 0.983$.
The bimodal log-normal distribution provides a left peak of the waiting time near 34 ms, which is quite similar to FRB 20201124A (39 ms in Ref.\cite{xu2022} and 51 ms in Ref.\cite{zhang2022}) and FRB 20220912A (51 ms in Ref.\cite{zhang2023}). The presence of closely spaced characteristic waiting times among repeating bursts may reflect intrinsic properties of the FRB source, such as the parameters of the aftershock activation function in the rotation-modulated starquake model\cite{luo2025}.
The right peak of the waiting time represents the activity of the FRB source during the statistical period. 

We also analyzed whether the distribution of waiting times has a fluence dependence. The distributions of waiting time in fluence regions of $F_\nu \leq$ 100 mJy~ms, 100 mJy~ms $ < F_\nu \leq$ 500 mJy~ms, and $F_\nu  > $ 500 mJy~ms were counted and analyzed separately. In Extended Data Fig.~\ref{exfig2}b, the left waiting time peaks are similar across the three groups (51.38 ms for $F_\nu \leq$ 100 mJy~ms, 52.93 ms for 100 mJy~ms  $< F_\nu \leq$ 500 mJy~ms, and 36.78 ms for $F_\nu>$ 500 mJy~ms) while the offset of the right peaks (22.86 s, 11.19 s, and 18.07 s, respectively) can be attributed to the differences in burst rates across the fluence ranges. Consequently, no significant variation is observed in the left peak of the waiting time distribution across different fluence domains, though a modulation appears in the right peak. Further periodicity analysis in the time domain is presented in an associated paper\cite{zhou2025}.

\begin{figure*}
\centering
\includegraphics[width=1\textwidth]{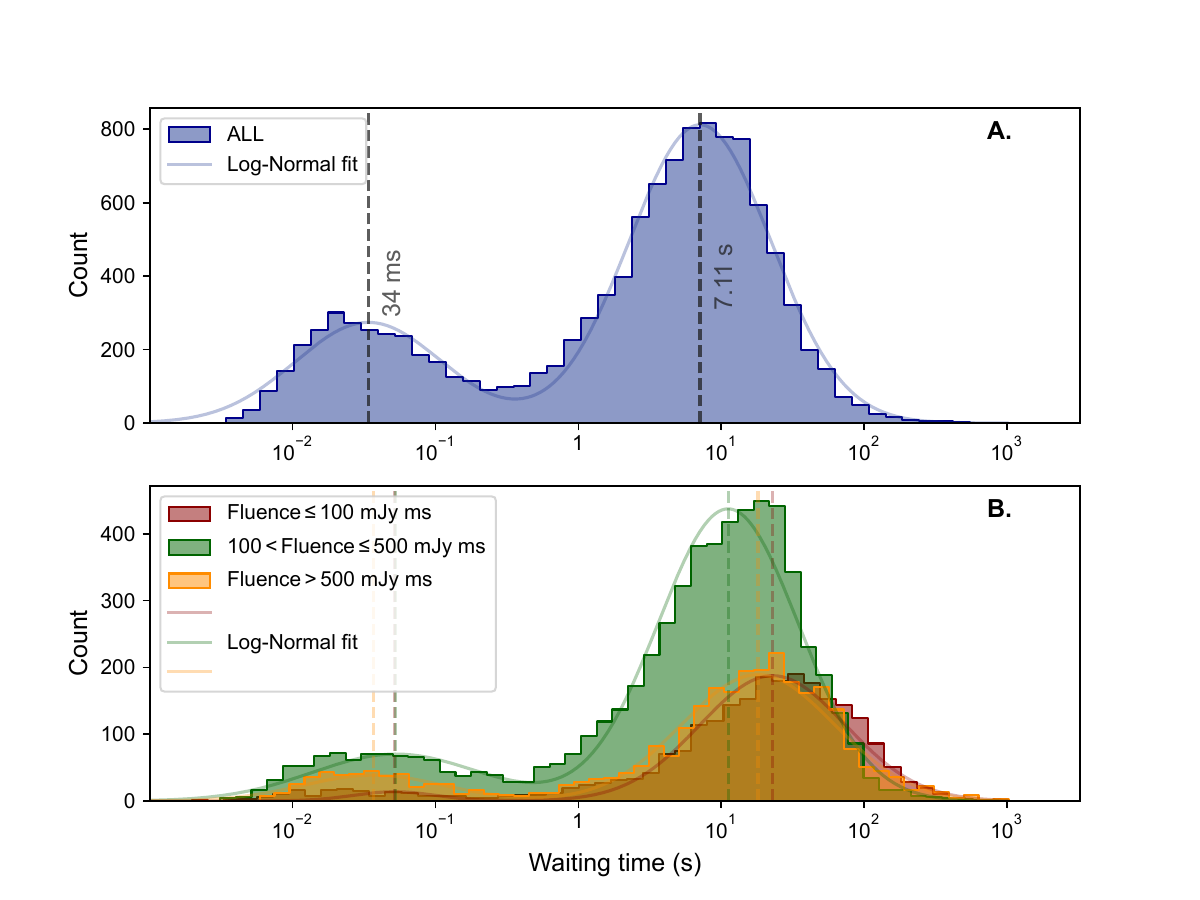}
\caption{{\bf The waiting time distribution of FRB 20240114A.} 
Panel A: The blue bars denote the waiting time histograms of all the bursts from FRB 20240114A. The blue solid line is the best-fit model with two log-normal functions. The dashed lines indicate the locations of maximum values in the two log-normal functions. The longer timescale peak is centered around 7.11 s, while the shorter timescale peak is at 34 ms.
Panel B: The waiting time distributions are shown for three distinct energy ranges. The red, green, and orange bars correspond to bursts with energies of $F_\nu \leq$ 100 mJy~ms, 100 mJy~ms$< F_\nu \leq$ 500 mJy~ms, and $F_\nu>$500 mJy~ms, respectively. For each case, the corresponding solid lines denote the best-fit models, and the corresponding dashed lines indicate the peak locations for the respective distributions.
}\label{exfig2}
\end{figure*}

\section{Comprehensive Analysis}

We present a detailed, comprehensive analysis on potential correlations among various measured parameters of FRB 20240114A (see Supplementary Table~\ref{tab:bursttab} for the full catalog) in this section. We investigate the potential correlation among various key parameter pairs of the bursts, which include the DM, central frequency, bandwidth, equivalent burst duration, peak flux density, and isotropic equivalent energy, aiming to gain further insights into the triggering and the emission mechanism of repeating FRBs. 

Extended Data Fig.~\ref{exfig3} shows the pair-wise correlation analysis between the measured parameters of FRB 20240114A. Notably, the bandwidth-central frequency plane displays a triangular shape, as the central frequency marks the midpoint of the observed frequency range. There is no obvious correlation among most parameter pairs, except for the peak flux density-isotropic equivalent energy, the bandwidth-isotropic equivalent energy, and the equivalent burst duration-isotropic equivalent energy planes, which arise from parameter coupling. Therefore, most of the parameters are incoherent with each other.

We calculate the PCC for each of the parameter pairs, which reveals that only the flux density-energy plane and the bandwidth-energy plane yields PCC values higher than 0.5, with values of 0.86 and 0.67, respectively.
The positive correlations can be attributed to the fact that the isotropic equivalent energy is proportional to the peak flux density and bandwidth. 
The $W_{\rm eq}$ distribution spans a relatively narrow range (within approximately one order of magnitude), whereas the peak flux density distribution extends across nearly five orders of magnitude.
As a result, the energy distribution is primarily dominated by the peak flux density, leading to a stronger positive correlation in the flux-energy plane.

\begin{figure*}
\centering
\includegraphics[width=1\textwidth]{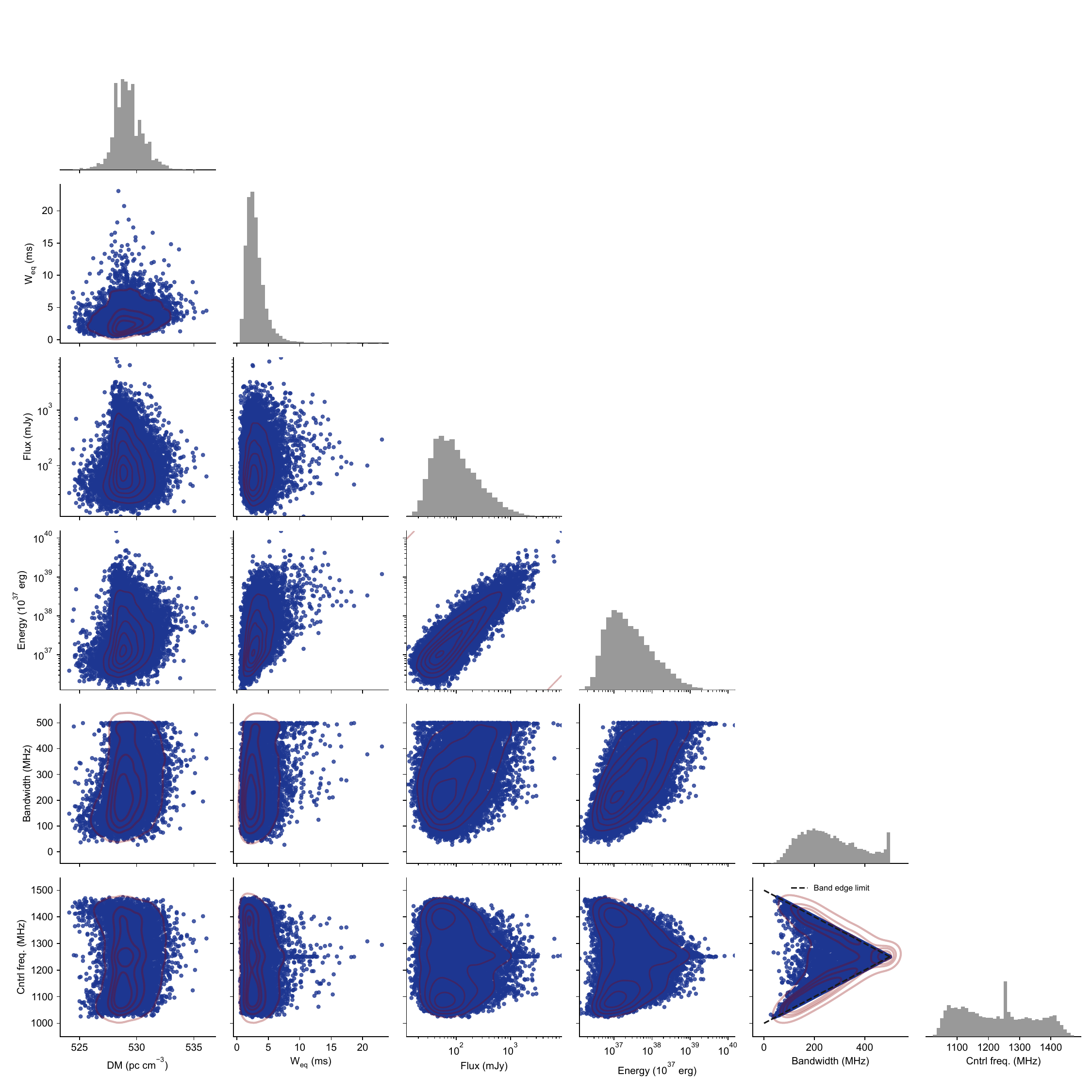}
\caption{{\bf The pairwise correlation analysis between the observation properties of the bursts.} The blue dots represent the parameters of each burst. The red contour lines present in each panel are the result of the KDE fitting. The gray histograms represent the parameter distributions of the bursts. The dashed black triangular shape in the Bandwidth-Central frequency panel is due to the limits of our observing band.}

\label{exfig3}

\end{figure*}
\clearpage
\bibliographystyle{naturemag}

\subsection{Data availability}
Observational properties of 11,553 burst events of FRB 20240114A measured with FAST from January 2024 to August 2024 are summarized in the manuscript Supplementary Table \ref{tab:bursttab} and  \url{https://doi.org/10.57760/sciencedb.Fastro.00030}. Observational data are available from the FAST archive\url{https://fast.bao.ac.cn}. Due to the large data volume for these observations, interested users are encouraged to contact the corresponding author to arrange the data transfer.

\subsection{Code availability}
Computational programs for the FRB 20240114A burst analysis and observations reported here are available at \url{https://github.com/NAOC-pulsar/PeiWang-code}. Other standard data reduction packages are available at their respective websites:\\
PRESTO: \url{https://github.com/scottransom/presto}\\
DRAFTS:  \url{https://github.com/SukiYume/DRAFTS} \\
DSPSR:    \url{http://dspsr.sourceforge.net}\\
PSRCHIVE:  \url{http://psrchive.sourceforge.net}\\
DM-power:  \url{https://github.com/hsiuhsil/DM-power}

\clearpage
\section*{Supplementary Table}
\renewcommand{\baselinestretch}{1.0}
\selectfont
\noindent
\EXTTAB{secA1}~1: Observations of FRB 20240114A from Jan. 2024 to Aug. 2024.\\
\EXTTAB{secA2}~2: The properties of 11,553 bursts of FRB 20240114A measured with FAST.\\
\EXTTAB{secA3}~3: The fitted parameters of the burst rate distribution of energy.\\
\EXTTAB{secA4}~4: The energy budgets of five repeating FRBs.

\setcounter{figure}{0}
\setcounter{table}{0}
\captionsetup[table]{name={\bf Supplementary Table} }
\setlength{\tabcolsep}{1mm}{
\renewcommand\arraystretch{1.2}
\scriptsize
\centering
\begin{longtable}{c c c c c c c c}
\caption{Observations of FRB 20240114A from Jan. 2024 to Aug. 2024.}\\
\hline\hline
Date UT & MJD$_{\rm start}^{a}$ & MJD$_{\rm end}^{b}$ & Freq. resolution & t$_{\rm samp}$ & Duration & Num. of & Burst Rate$^{c}$ \\%
($\mathrm{YYYYMMDD}$)& (at infinite freq.)& (at infinite freq.) & (kHz) & ($\mu$s) & ($\mathrm{hour}$) & Detections  & ($\mathrm{hour^{-1}}$) \\%
\hline
\endfirsthead
\hline\hline
Date UT & MJD$_{\rm start}^{a}$ & MJD$_{\rm end}^{b}$ & Freq. resolution & t$_{\rm samp}$ & Duration & Num. of & Burst Rate$^{c}$ \\%
($\mathrm{YYYYMMDD}$)&(at infinite freq.) & (at infinite freq.) & (kHz) & ($\mu$s) & ($\mathrm{hour}$) & Detections  & ($\mathrm{hour^{-1}}$) \\%
\hline
\endhead
\hline
\endfoot
\hline
\endlastfoot

20240128 & 60337.229849817 & 60337.250683150 & 122.070 & 196.608 & 0.50 & 8 & 16.0\\ 
20240129 & 60338.215266484 & 60338.236099817 & 122.070 & 49.152 & 0.50 & 11 & 22.0\\ 
20240201 & 60341.179849817 & 60341.200683150 & 122.070 & 49.152 & 0.50 & 9 & 18.0\\ 
20240204 & 60344.240960928 & 60344.261794262 & 122.070 & 49.152 & 0.50 & 2 & 4.0\\ 
20240209 & 60349.152766484 & 60349.173599817 & 122.070 & 49.152 & 0.50 & 6 & 12.0\\ 
20240213 & 60353.215960928 & 60353.236794262 & 122.070 & 49.152 & 0.50 & 14 & 28.0\\ 
20240215 & 60355.131238706 & 60355.152072039 & 122.070 & 49.152 & 0.50 & 21 & 42.0\\ 
20240221 & 60361.179155373 & 60361.199988706 & 122.070 & 49.152 & 0.50 & 24 & 48.0\\ 
20240225 & 60365.131933150 & 60365.152766484 & 122.070 & 49.152 & 0.50 & 11 & 22.0\\ 
20240305 & 60374.188183150 & 60374.209016484 & 61.035 & 98.304 & 0.50 & 225 & 450.0\\ 
20240306 & 60375.134710928 & 60375.218044262 & 122.070 & 49.152 & 2.00 & 918 & 459.0\\ 
20240307 & 60376.102766484 & 60376.186099817 & 122.070 & 49.152 & 2.00 & 969 & 484.5\\ 
20240308 & 60377.193044262 & 60377.221447039 & 122.070 & 49.152 & 0.68 & 347 & 509.0\\ 
20240309 & 60378.191655373 & 60378.205660002 & 122.070 & 49.152 & 0.34 & 186 & 553.4\\ 
20240310 & 60379.199294262 & 60379.219120650 & 122.070 & 49.152 & 0.48 & 270 & 567.4\\ 
20240311 & 60380.152766484 & 60380.194433150 & 122.070 & 49.152 & 1.00 & 727 & 727.0\\ 
20240312 & 60381.038183150 & 60381.220822039 & 122.070 & 49.152 & 4.38 & 3197 & 729.4\\ 
20240313 & 60382.173599817 & 60382.215266484 & 122.070 & 49.152 & 1.00 & 535 & 535.0\\ 
20240315 & 60384.190960928 & 60384.204849817 & 122.070 & 49.152 & 0.33 & 66 & 198.0\\ 
20240317 & 60386.118738706 & 60386.139572039 & 122.070 & 49.152 & 0.50 & 171 & 342.0\\ 
20240321 & 60390.158322039 & 60390.172210928 & 61.035 & 98.304 & 0.33 & 111 & 333.0\\ 
20240327 & 60396.133322039 & 60396.154155373 & 122.070 & 49.152 & 0.50 & 297 & 594.0\\ 
20240329 & 60398.981933150 & 60399.002766484 & 122.070 & 49.152 & 0.50 & 275 & 550.0\\ 
20240401 & 60401.066655373 & 60401.087488706 & 122.070 & 49.152 & 0.50 & 261 & 522.0\\ 
20240403 & 60403.046516484 & 60403.067349817 & 122.070 & 49.152 & 0.50 & 141 & 282.0\\ 
20240407 & 60407.091655373 & 60407.112488706 & 122.070 & 49.152 & 0.50 & 41 & 82.0\\ 
20240410 & 60410.020822039 & 60410.041655373 & 122.070 & 49.152 & 0.50 & 35 & 70.0\\ 
20240413 & 60413.023599817 & 60413.044433150 & 122.070 & 49.152 & 0.50 & 44 & 88.0\\ 
20240417 & 60417.078460928 & 60417.099294262 & 122.070 & 49.152 & 0.50 & 45 & 90.0\\ 
20240421 & 60421.009016484 & 60421.029849817 & 122.070 & 49.152 & 0.50 & 72 & 144.0\\ 
20240424 & 60424.983322039 & 60425.004155373 & 122.070 & 49.152 & 0.50 & 110 & 220.0\\ 
20240430 & 60430.029155373 & 60430.049988706 & 122.070 & 49.152 & 0.50 & 133 & 266.0\\ 
20240504 & 60434.956933150 & 60434.977766484 & 122.070 & 49.152 & 0.50 & 89 & 178.0\\ 
20240511 & 60441.888877595 & 60441.909710928 & 122.070 & 49.152 & 0.50 & 102 & 204.0\\ 
20240518 & 60448.897210928 & 60448.911099817 & 122.070 & 49.152 & 0.33 & 32 & 96.0\\ 
20240525 & 60455.913877595 & 60455.927766484 & 122.070 & 49.152 & 0.33 & 37 & 111.0\\ 
20240531 & 60461.949294262 & 60461.963183150 & 122.070 & 49.152 & 0.33 & 11 & 33.0\\ 
20240608 & 60469.827766484 & 60469.841655373 & 122.070 & 49.152 & 0.33 & 49 & 147.0\\ 
20240615 & 60476.877072039 & 60476.890960928 & 122.070 & 49.152 & 0.33 & 29 & 87.0\\ 
20240622 & 60483.881933150 & 60483.895822039 & 122.070 & 49.152 & 0.33 & 40 & 120.0\\ 
20240629 & 60490.862488706 & 60490.876377595 & 122.070 & 49.152 & 0.33 & 46 & 138.0\\ 
20240706 & 60497.858322039 & 60497.872210928 & 122.070 & 49.152 & 0.33 & 42 & 126.0\\ 
20240713 & 60504.761099817 & 60504.774988706 & 122.070 & 49.152 & 0.33 & 117 & 351.0\\ 
20240720 & 60511.684710928 & 60511.698599817 & 122.070 & 49.152 & 0.33 & 100 & 300.0\\ 
20240722 & 60513.800683150 & 60513.815266484 & 122.070 & 49.152 & 0.35 & 92 & 262.9\\ 
20240724 & 60515.773599817 & 60515.834710928 & 61.035 & 98.304 & 1.47 & 369 & 251.6\\ 
20240726 & 60517.819433150 & 60517.840266484 & 122.070 & 49.152 & 0.50 & 166 & 332.0\\ 
20240727 & 60518.742349817 & 60518.756238706 & 122.070 & 49.152 & 0.33 & 92 & 276.0\\ 
20240728 & 60519.765266484 & 60519.786099817 & 122.070 & 49.152 & 0.50 & 208 & 416.0\\ 
20240730 & 60521.799988706 & 60521.820822039 & 122.070 & 49.152 & 0.50 & 138 & 276.0\\ 
20240802 & 60524.783322039 & 60524.797210928 & 122.070 & 49.152 & 0.33 & 108 & 324.0\\ 
20240805 & 60527.777766484 & 60527.791655373 & 122.070 & 49.152 & 0.33 & 52 & 156.0\\ 
20240808 & 60530.774988706 & 60530.788877595 & 122.070 & 49.152 & 0.33 & 90 & 270.0\\ 
20240814 & 60536.777766484 & 60536.791655373 & 122.070 & 49.152 & 0.33 & 86 & 258.0\\ 
20240817 & 60539.709710928 & 60539.723599817 & 122.070 & 49.152 & 0.33 & 70 & 210.0\\ 
20240821 & 60543.737488706 & 60543.751377595 & 122.070 & 49.152 & 0.33 & 66 & 198.0\\ 
20240829 & 60551.699294262 & 60551.713183150 & 122.070 & 49.152 & 0.33 & 40 & 120.0\\ 
\label{tab:obslog}
\end{longtable}}
\begin{tablenotes}
\item[a] $^a$ The start time of each observation session with topocentric coordinates, corrected to infinite frequency.
\item[b] $^b$ The end time of each observation session with topocentric coordinates, corrected to infinite frequency.
\item[c] $^c$ The burst rate is assumed to be a constant throughout the day, defined as the number of detections divided by the duration of observation. 
\end{tablenotes}

\clearpage
\setcounter{figure}{1}
\setcounter{table}{1}
\captionsetup[table]{name={\bf Supplementary Table} }
\setlength{\tabcolsep}{0.8mm}{
\renewcommand\arraystretch{1.1}
\scriptsize
\centering
\begin{longtable}{c c c c c c c c c c c c}%
\caption{The properties of 11,553 bursts of FRB 20240114A measured with FAST$^a$.}\\
\hline\hline%
Burst & MJD $_{\rm bary}^{a}$ & DM$^{b}$ & W$_{\rm eq}^{c}$ & Cntrl freq.$^{d}$ & Bandwidth$^{e}$ & Peak\ flux$^{f}$ & Fluence & Energy$_{\Delta\nu}^{g)}$ \\%
ID&(at infinite freq.) & ($\mathrm{pc\ cm^{-3}}$) & (ms)  & (MHz) & (MHz) & (mJy) & (mJy\ ms)  & ($\mathrm{\times 10^{37}\ erg}$)  \\%
\hline%
\endhead%
\hline%
\endfoot%
\hline%
\endlastfoot%
B00001 & 60337.233896015 & 529.5$\pm$0.1 & 3.51$\pm$0.23 & 1105 & 188$\pm$11$^{3)}$ & 217.5$\pm$9.6 & 764$\pm$61 & 6.12$\pm$0.60  \\ 
B00002 & 60337.235854917 & 529.6$\pm$0.8 & 2.60$\pm$0.43 & 1041 & 75$\pm$8$^{2)}$ & 86.5$\pm$3.8 & 225$\pm$39 & 0.72$\pm$0.15  \\ 
B00003 & 60337.237146105 & 525.0$\pm$0.3 & 4.63$\pm$1.23 & 1167 & 355$\pm$67$^{3)}$ & 18.1$\pm$0.9 & 84$\pm$23 & 1.27$\pm$0.42  \\ 
B00004 & 60337.239375175 & 526.7$\pm$2.0 & 2.66$\pm$0.49 & 1039 & 74$\pm$5$^{2)}$ & 74.3$\pm$3.3 & 198$\pm$38 & 0.62$\pm$0.13  \\ 
B00005 & 60337.239452224 & 527.0$\pm$2.2 & 1.20$\pm$0.26 & 1064 & 104$\pm$4$^{3)}$ & 77.4$\pm$3.6 & 93$\pm$21 & 0.41$\pm$0.09  \\ 
B00006 & 60337.243643750 & 527.7$\pm$0.2 & 3.67$\pm$0.38 & 1058 & 111$\pm$11$^{2)}$ & 127.5$\pm$4.9 & 468$\pm$52 & 2.20$\pm$0.33  \\ 
B00007 & 60337.245743427 & 527.7$\pm$0.1 & 3.02$\pm$0.21 & 1106 & 208$\pm$6$^{2)}$ & 257.9$\pm$11.6 & 780$\pm$65 & 6.88$\pm$0.61  \\ 
B00008 & 60337.246124366 & 526.8$\pm$0.3 & 3.72$\pm$0.53 & 1058 & 99$\pm$14$^{3)}$ & 98.2$\pm$4.5 & 365$\pm$55 & 1.53$\pm$0.32  \\ 
B00009 & 60338.212964888 & 528.5$\pm$0.1 & 1.62$\pm$0.06 & 1291 & 319$\pm$8$^{2)}$ & 224.6$\pm$7.9 & 363$\pm$19 & 4.93$\pm$0.29  \\ 
B00010 & 60338.216853487 & 527.9$\pm$0.1 & 5.75$\pm$1.02 & 1100 & 228$\pm$37$^{3)}$ & 37.8$\pm$0.8 & 217$\pm$39 & 2.10$\pm$0.51  \\ 

...    & ...              & ... & ...           & ...  & ...       & ...          & ...            & ... \\
B11543 & 60551.715922689 & 528.6$\pm$0.2 & 5.51$\pm$0.80 & 1392 & 389$\pm$23$^{3)}$ & 30.8$\pm$2.9 & 170$\pm$29 & 2.80$\pm$0.52  \\ 
B11544 & 60551.716676723 & 530.4$\pm$0.1 & 2.02$\pm$0.11 & 1220 & 407$\pm$11$^{2)}$ & 82.9$\pm$7.6 & 167$\pm$18 & 2.90$\pm$0.32  \\ 
B11545 & 60551.716689314 & 528.4$\pm$0.3 & 2.41$\pm$0.23 & 1395 & 171$\pm$16$^{2)}$ & 69.1$\pm$6.0 & 167$\pm$21 & 1.21$\pm$0.19  \\ 
B11546 & 60551.716691701 & 532.1$\pm$0.1 & 9.38$\pm$0.37 & 1144 & 249$\pm$50$^{1)}$ & 115.2$\pm$10.3 & 1081$\pm$106 & 11.45$\pm$2.55  \\ 
B11547 & 60551.716691923 & 533.0$\pm$0.3 & 8.35$\pm$0.32 & 1133 & 243$\pm$5$^{2)}$ & 132.2$\pm$11.5 & 1104$\pm$105 & 11.39$\pm$1.11  \\ 
B11548 & 60551.717399910 & 529.2$\pm$1.6 & 6.99$\pm$1.65 & 1206 & 198$\pm$6$^{3)}$ & 17.5$\pm$1.5 & 122$\pm$31 & 1.03$\pm$0.26  \\ 
B11549 & 60551.718067969 & 533.2$\pm$0.7 & 3.22$\pm$0.35 & 1274 & 172$\pm$5$^{3)}$ & 42.1$\pm$3.5 & 135$\pm$19 & 0.99$\pm$0.14  \\ 
B11550 & 60551.718584716 & 529.3$\pm$0.5 & 2.55$\pm$0.52 & 1251 & 480$\pm$25$^{3)}$ & 17.9$\pm$1.6 & 46$\pm$10 & 0.94$\pm$0.21  \\ 
B11551 & 60551.718678457 & 528.3$\pm$1.5 & 2.53$\pm$0.54 & 1071 & 116$\pm$7$^{2)}$ & 29.9$\pm$2.6 & 76$\pm$17 & 0.37$\pm$0.09  \\ 
B11552 & 60551.718785581 & 533.2$\pm$0.2 & 5.73$\pm$0.39 & 1194 & 340$\pm$11$^{2)}$ & 64.0$\pm$6.0 & 366$\pm$42 & 5.30$\pm$0.63  \\ 
B11553 & 60551.718925823 & 530.2$\pm$1.4 & 4.34$\pm$0.84 & 1218 & 394$\pm$79$^{1)}$ & 20.1$\pm$1.8 & 87$\pm$18 & 1.46$\pm$0.42  \\

\label{tab:bursttab}
\end{longtable}}
\begin{tablenotes}
\item[a)] $^a$ The full table is available in ScienceDB, doi: \url{https://doi.org/10.57760/sciencedb.Fastro.00030}.\\
\item[b)] $^b$ Arrival time of burst peak at the solar system barycenter, after correcting to the infinite frequency. \\%
\item[c)] $^c$ DM obtained from the best burst alignment. A '-' is marked and the daily averaged DM value is used where the DM-fit failed.\\%
\item[d)] $^d$ The equivalent width W$_{\rm eq}$ was defined as the width of a rectangular burst that has the same area as the profiles, with the height of peak flux density. \\%
\item[e)] $^e$ The central frequency is defined as the median value of the frequency range of the burst.\\
\item[f)] $^f$ Three types of burst bandwidth fit were considered: $^{\#}$1) Boxcar bandwidth of the burst with complex morphological
structures or those that were too faint. A conservative 20$\%$ fractional error is assumed; $^{\#}$2) CDF frequency boundaries of the bursts, the uncertainty of measurement was estimated using the bootstrap method; $^{\#}$3) A frequency extended Gaussian fit was used to correct the bandwidth for the bursts' cut off by the upper/lower limit of the observed frequency, 1$\sigma$  measuring uncertainty. \\%

\item[g)] $^g$ The average peak flux density over their individual frequency range of bursts. A '*' indicates bursts affected by saturation, which may lead to a potential underestimation of their flux density. \\%
\item[h)] $^h$ The isotropic energy of the bursts was calculated using a measured frequency bandwidth, see text. \\%
\end{tablenotes}
\clearpage
\setcounter{figure}{2}
\setcounter{table}{2}
\captionsetup[table]{name={\bf Supplementary Table} }
\setlength{\tabcolsep}{4mm}{
\renewcommand\arraystretch{1.5}
%\footnotesize 
\scriptsize
% \tiny
\centering
\begin{longtable}{c c c c c}
\caption{The fitted parameters of the burst rate distribution of energy} \label{tab:fitting} \\
\hline\hline
Model & Fitting parameter & $\bar{R}^2$ & Reduced $\chi^2$ & AIC \\
\hline
\endfirsthead
\hline\hline
Model & Fitting parameter & $\bar{R}^2$ & Reduced $\chi^2$ & AIC \\
\hline
\endhead
\hline
\endfoot
\hline
\endlastfoot
  LN     & $E_{c~}= 1.71\times10^{37}$ erg, $\sigma_{~}=0.57$ & 0.918 & 13.38 & 58.00 \\
  Bi-LN     & $E_{c1}= 1.00\times10^{37}$ erg, $\sigma_1=0.56$  & 0.986 & 2.38 & 13.64 \\
     & $E_{c2}= 3.87\times10^{37}$ erg, $\sigma_2=0.31$  & & &\\

  % LN     & $E_{c~}= 1.71\pm0.11\times10^{37}$ erg, $\sigma_{~}=0.57\pm0.03$ & 0.918 & 13.38 & 58.00 \\
  % Bi-LN     & $E_{c1}= 1.00\pm0.05\times10^{37}$ erg, $\sigma_1=0.56\pm 0.06$  & 0.986 & 2.38 & 13.64 \\
     % & $E_{c2}= 3.87\pm0.17\times10^{37}$ erg, $\sigma_2=0.31\pm0.13 $  & & &\\

\end{longtable}
}

\clearpage
\setcounter{figure}{3}
\setcounter{table}{3}
\captionsetup[table]{name={\bf Supplementary Table} }
\setlength{\tabcolsep}{2mm}{
\renewcommand\arraystretch{1.5}
%\footnotesize 
\scriptsize
% \tiny
\centering
\begin{longtable}{c c c c c}
\caption{The energy budgets of five repeating FRBs.} \label{tab:energycal} \\
\hline\hline
FRB Name & Duty Cycle$^a$ & Total Radio Energy$^b$ & Averaged Energy$^c$ & Source Energy$^d$ \\
         & $\zeta$        &  $E_{\rm tot}$(erg)      & $E_{\rm avg}$(erg hr$^{-1}$) & $E_{\rm src}$($\eta_{r,-4}^{-1}F_{b,-1}$ erg) \\
\hline
\endfirsthead
\hline\hline
FRB Name & Duty Cycle$^a$ & Radio E$^b$ & Averaged E$^c$ & Source E$^d$ \\
         & $\zeta$        &  (erg)      & (erg hr$^{-1}$) &       ($\eta_{r,-4}^{-1}F_{b,-1}$ erg) \\
\hline
\endhead
\hline
\endfoot
\hline
\endlastfoot

  20121102A     & 0.053  & $1.36\times10^{41}$ & $2.29\times10^{39}$ & $2.59\times10^{45}$ \\
  20190520B     & 0.070  & $4.39\times10^{39}$ & $2.37\times10^{38}$ & $6.26\times10^{43}$ \\
  20201124A$^{e}$ & 0.063  & $1.65\times10^{41}$ & $2.01\times10^{39}$ & $2.60\times10^{45}$ \\
  20201124A$^f$ & 0.042  & $6.42\times10^{40}$ & $1.60\times10^{40}$ & $1.54\times10^{45}$ \\
  20220912A     & 0.021  & $7.42\times10^{40}$ & $8.55\times10^{39}$ & $3.49\times10^{45}$ \\
  20240114A     & 0.0066  & $9.62\times10^{41}$ & $2.84\times10^{40}$ & $1.47\times10^{47}$ \\

\end{longtable}
}
\begin{tablenotes}
\item[a] $^a$ The observation duty cycle, e.g., for FRB~20240114A in this paper, the duty cycle is 33.86 hours out of 214 days.
\item[b] $^b$ Sum of the observed isotropic radio energies of all bursts. 
\item[c] $^c$ The total radio energy divided by observation time, e.g. for FRB~20240114A in this paper, the averaged energy is $9.62\times10^{41}$ erg / 33.86 hours. 
\item[d] $^d$ The total source energy calculated with Eq. \ref{eq:cal1},  using $\eta_{\rm r}=10^{-4}\eta_{\rm r,-4}^{-1}$ and $F_{\rm b}=0.1F_{\rm b,-1}$. 
\item[e] $^e$ FAST observation of FRB~20201124A in 2021.04 by Ref.\cite{xu2022}
\item[f] $^f$ FAST observation of FRB~20201124A in 2021.09 by Ref.\cite{zhang2022}

\end{tablenotes}
\end{methods}
\end{document}